\documentclass[a4paper,11pt]{article}
\pdfoutput=1
\usepackage{units}
\usepackage{amstext}
\usepackage{graphicx}

\makeatletter
\usepackage{pos}

\title{Jet substructure in $p$+$p$ and $p$+Au collisions at $\sqrt{s_{\rm{NN}}}=200$ GeV at STAR}
\ShortTitle{Jet substructure at STAR}

\author*[a]{Isaac Mooney for the STAR Collaboration}

\affiliation[a]{Wayne State University,\\
  666 W. Hancock St., Detroit, USA}

\emailAdd{fs3383@wayne.edu}

\abstract{In order to attribute the partonic energy loss experienced by jets (jet quenching)
observed in A+A collisions to the traversal of partons through the
hot QCD medium, it is necessary to examine the cold nuclear matter (CNM)
effects on the corresponding jets. Such an examination has historically
been done using $p$+A collisions. We present fully corrected
measurements of the jet mass and SoftDrop groomed jet mass in $p$+$p$ and $p$+Au collisions at STAR at $\sqrt{s_{\rm{NN}}}=200$ GeV as a function of the event activity (EA) to increase or decrease the
magnitude of CNM effects. EA is determined in the backward (Au-going)
rapidity ($-5.0<\eta<-3.3$) by the STAR Beam-Beam Counter to minimize auto-correlation with jets measured
at mid-rapidity. Comparison of the jet mass distribution in $p$+Au collisions to that in $p$+$p$ collisions allows for isolation of CNM effects in anticipation of an upcoming jet mass measurement in Au+Au collisions.}

\FullConference{%
  HardProbes2020\\
  1-6 June 2020\\
  Austin, Texas}

\makeatother

\begin{document}
\maketitle

\section{Introduction\label{sec:Introduction}}

\addtocounter{page}{-1}

\thispagestyle{plain}

As hard partons from high-$Q^{2}$ scatterings evolve in vacuum, they
radiate stochastically and fragment into hadrons which lend the resulting
jet a unique substructure. Studying jet substructure gives insight
into various aspects of QCD, from the initial hard scattering, to
the parton shower, and eventual hadronization. Jet mass, $M$, is
one such substructure observable, belonging to a generic class of
angularity observables \cite{Kangularity}, defined as the magnitude
of the four-momentum sum of constituents ($M=|\sum_{i\in J}p_{i}|=\sqrt{E^{2}-\mathbf{p}^{2}}$).
The mass of a reconstructed jet is a proxy for the initial parton's
virtuality \cite{MajumderPutschke}. Measurements of the jet mass
are expected to provide vital inputs to Monte Carlo (MC) models' implementations
of parton shower and hadronization algorithms. In addition, to focus
on the perturbative parton shower, we suppress wide-angle non-perturbative
(soft) radiation with SoftDrop grooming \cite{SoftDrop} and report
both groomed and ungroomed jet mass.

In 2016, the PHENIX collaboration reported an unexpected jet production
enhancement (suppression) in peripheral (central) $p$+Au collisions
compared to $p$+$p$ collisions \cite{PHENIX}. Studying jet substructure
in $p$+Au collisions will help determine whether this effect is due
to jet modification in a cold nuclear medium. Addressing this question
is necessary for interpreting measurements of jet mass in a hot nuclear
medium.

\section{Measurement\label{sec:Measurement}}

This study utilizes STAR data for proton-proton collisions at $\sqrt{s}=\unit[200]{GeV}$
from 2012 and proton-gold collisions at $\sqrt{s_{{\rm NN}}}=\unit[200]{GeV}$
from 2015. For both collision systems, we require a jet patch trigger
(see \cite{Raghav}) and reconstruct jets from charged tracks in the
time projection chamber (TPC) and energy deposits in the barrel electromagnetic
calorimeter (BEMC) using the anti-$k_{\text{T}}$ algorithm with a
jet resolution parameter $R=0.4$. Event, track, tower, and jet selections
are the same as in \cite{Raghav} but for one additional selection:
we require jet mass above $\unit[1]{GeV}/c^{2}$, due to poor detector
resolution below this. We report jets with transverse momentum ($p_{\text{T}}$)
between $20$ and $\unit[45]{GeV}/c$.

To correct for detector effects such as tracking efficiency and momentum
resolution, we perform a two-dimensional ($M$, $p_{\text{T}}$) iterative
Bayesian unfolding implemented in the RooUnfold package \cite{RooUnfold}.
In the case of $p$+$p$ collisions, we construct a response matrix
with particle-level events simulated by PYTHIA-6.4.28 Perugia 2012
(a STAR tune) \cite{Perugia2012} and detector-level events simulated
by the PYTHIA events run through a GEANT-3 STAR detector simulation,
and embedded in $p$+$p$ zero-bias events as an estimate of background.
In $p$+Au collisions, we use the same particle-level events, but
embed these detector-level events further into a $p$+Au minimum-bias
background.

Systematic uncertainties are made up of four main components: a tracking
efficiency uncertainty of 4\%; a tower gain uncertainty of 3.8\%;
a hadronic correction variation from the nominal 100\% subtraction
of matched tracks' momenta from tower energy to 50\%; and uncertainties
on the unfolding procedure such as variation in iteration parameter
and the shape of the priors. Uncertainty in background estimation
for $p$+Au unfolding is also considered.

For $p$+Au collisions, the event activity (EA) is determined by deposited
energy in STAR's backward ($-5<\eta<-3.4$) inner Beam-Beam Counter
(iBBC) on the Au-going side of the detector. In this work, we compare
p+Au events with 0-50\% EA (high-EA) and 50-100\% EA (low-EA). These
wide ranges will be refined in future work.

\section{Results\label{sec:Results}}

\begin{figure}
\begin{centering}
\includegraphics[width=0.69\textwidth]{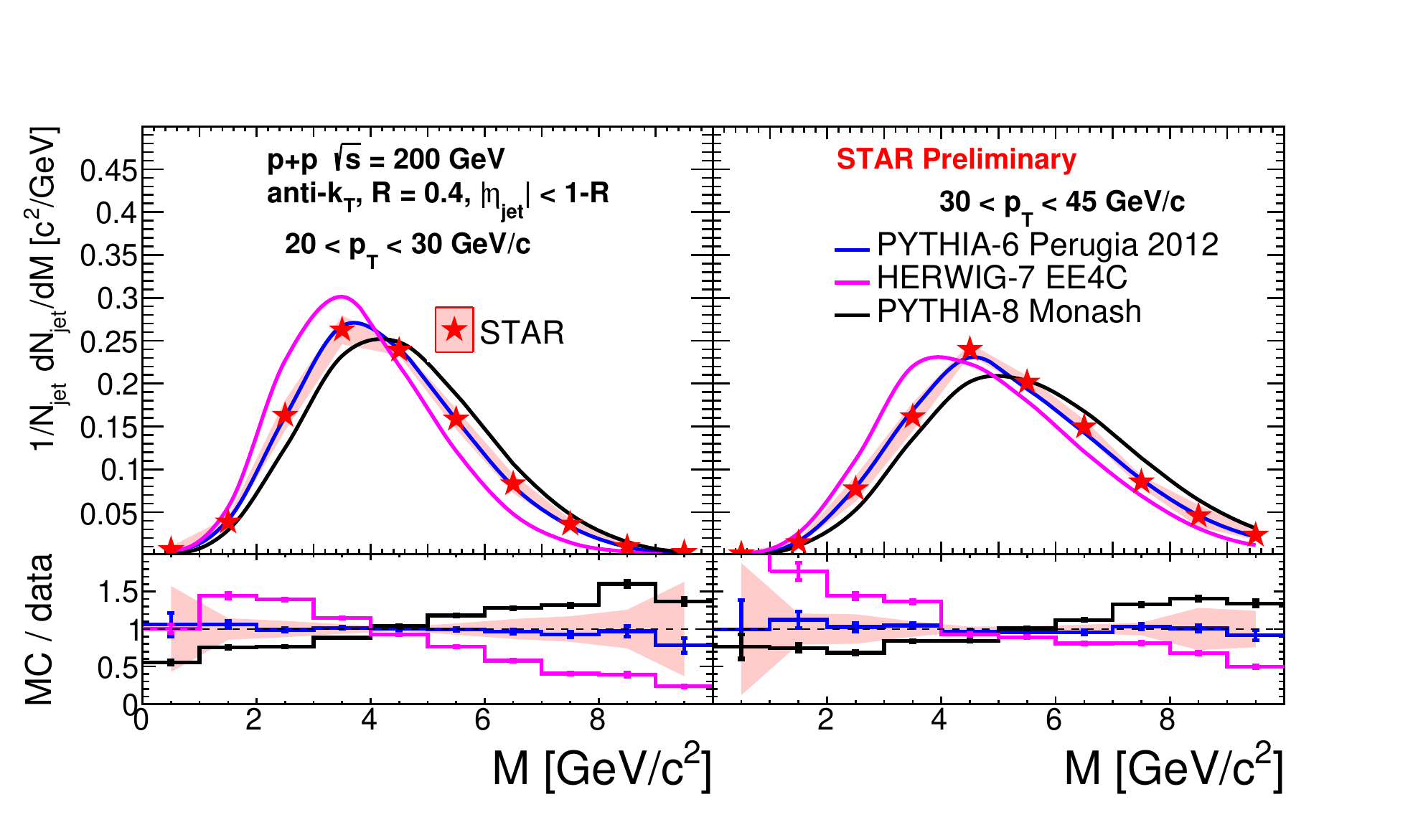}
\par\end{centering}
\caption{Measurement of the ungroomed jet mass, $M$, of anti-$k_{\text{T}}$
jets in $p$+$p$ collisions at $\sqrt{s}=\unit[200]{GeV}$ for two
jet $p_{\text{T}}$ ranges: $20<p_{\text{T}}<\unit[30]{GeV}/c$ (left),
and $30<p_{\text{T}}<\unit[45]{GeV}/c$ (right). The fully corrected
data (with red shaded band denoting systematic uncertainties) are
shown in solid red star markers. We compare via ratio to PYTHIA-6
(Perugia 2012 Tune, solid blue line), PYTHIA-8 (Monash Tune, solid
black line), and HERWIG-7 (EE4C Tune, solid magenta line) predictions.
In the lower panels the relative systematic uncertainty is drawn.
Statistical uncertainties are smaller than the size of the marker
in all figures.\label{fig:ppmass}}
\end{figure}
The fully corrected jet mass is shown for $p$+$p$ collisions in
Fig.~\ref{fig:ppmass} for $R=0.4$ jets with $20<p_{\text{T}}<\unit[30]{GeV}/c$
(left) and $30<p_{\text{T}}<\unit[45]{GeV}/c$ (right). As the jet
$p_{\text{T}}$ increases, we observe an increase in the mean jet
mass, as expected from pQCD, as well as a broadening of the distribution
due to the increase in the available phase space. We also compare
the results to three leading-order (LO) MC models: PYTHIA-6 with Perugia
2012 tune, PYTHIA-8 with Monash tune, and HERWIG-7 with EE4C tune,
where the latter two are tuned to LHC data. Relevant differences between
PYTHIA and HERWIG lie in the shower and hadronization mechanisms,
with PYTHIA using a $p_{\text{T}}$-ordered shower and string fragmentation,
while HERWIG utilizes an angular-ordered shower and cluster hadronization.
We note that PYTHIA-6 describes the data well within systematic uncertainties,
while the HERWIG-7 and PYTHIA-8 prefer lower and higher mass jets,
respectively.

\begin{figure}
\begin{centering}
\includegraphics[width=0.69\textwidth]{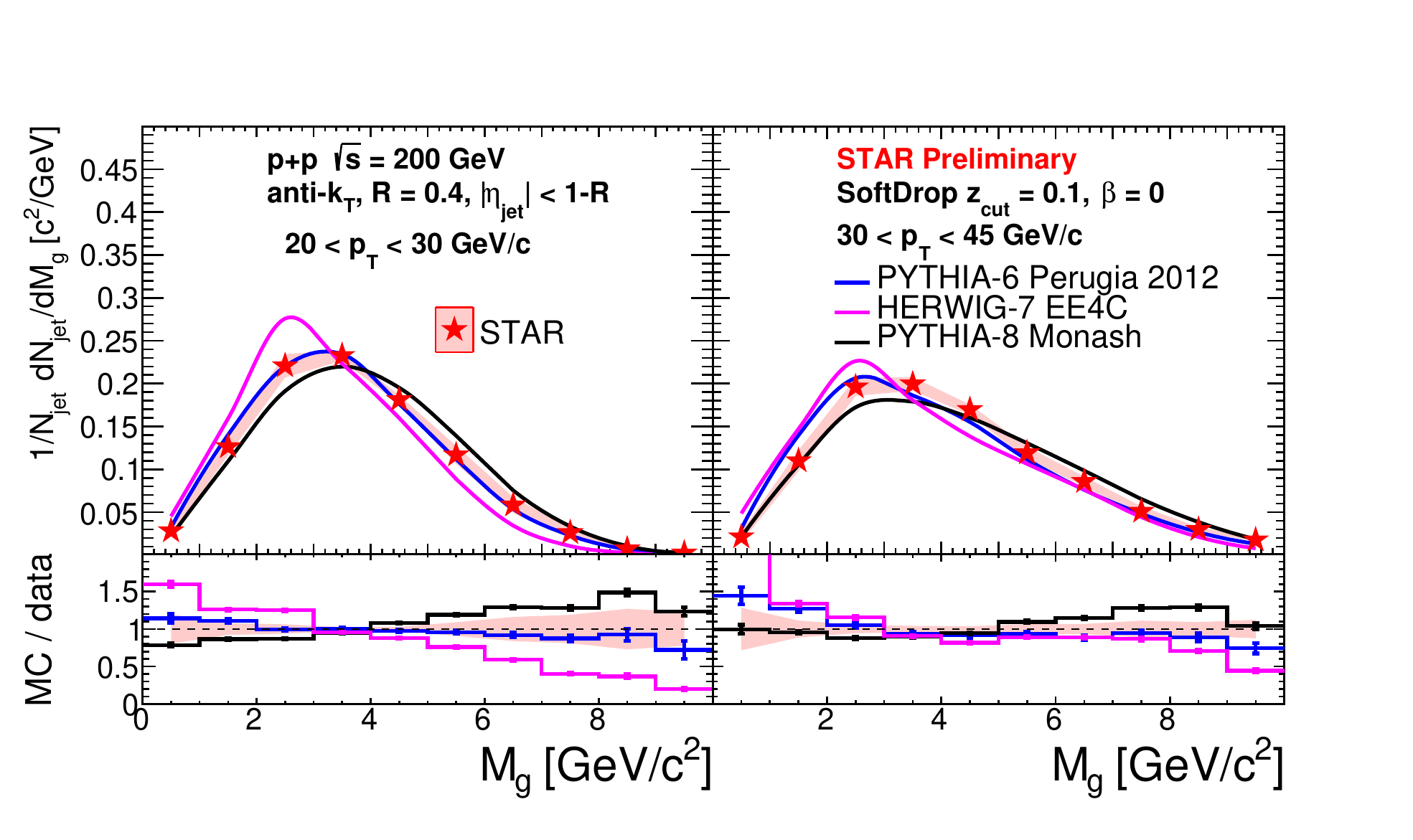}
\par\end{centering}
\caption{Measurement of the groomed mass, $M_{\text{g}}$, of anti-$k_{\text{T}}$
jets in $p$+$p$ collisions at $\sqrt{s}=\unit[200]{GeV}$. See Fig.~\ref{fig:ppmass}
for a description of the curves.\label{fig:ppmg}}
\end{figure}
Next, in order to remove jet constituents arising from soft radiation,
we apply the SoftDrop grooming algorithm in tagging mode with $z_{\text{cut}}=0.1$,
$\beta=0$. We report the groomed mass ($M_{\text{g}}$) in Fig.~\ref{fig:ppmg}
for ranges of the corresponding \emph{ungroomed} jet $p_{\text{T}}$
to allow direct comparison to the ungroomed jet mass. Note that $\left\langle M_{\text{g}}\right\rangle $
is reduced (\emph{cf}. Fig.~\ref{fig:ppmass}) due to the suppression
of non-perturbative effects. As before, we compare data (red star
markers) to the three LO MC models (solid lines) mentioned above,
in the ratio panel. Here we see a reduced systematic uncertainty on
the groomed jet mass. The trends are similar to the ungroomed jet
mass although disagreement between data and the two LHC-tuned MC models
(HERWIG-7 and PYTHIA-8) is reduced.

\begin{figure}
\begin{centering}
\includegraphics[width=0.69\textwidth]{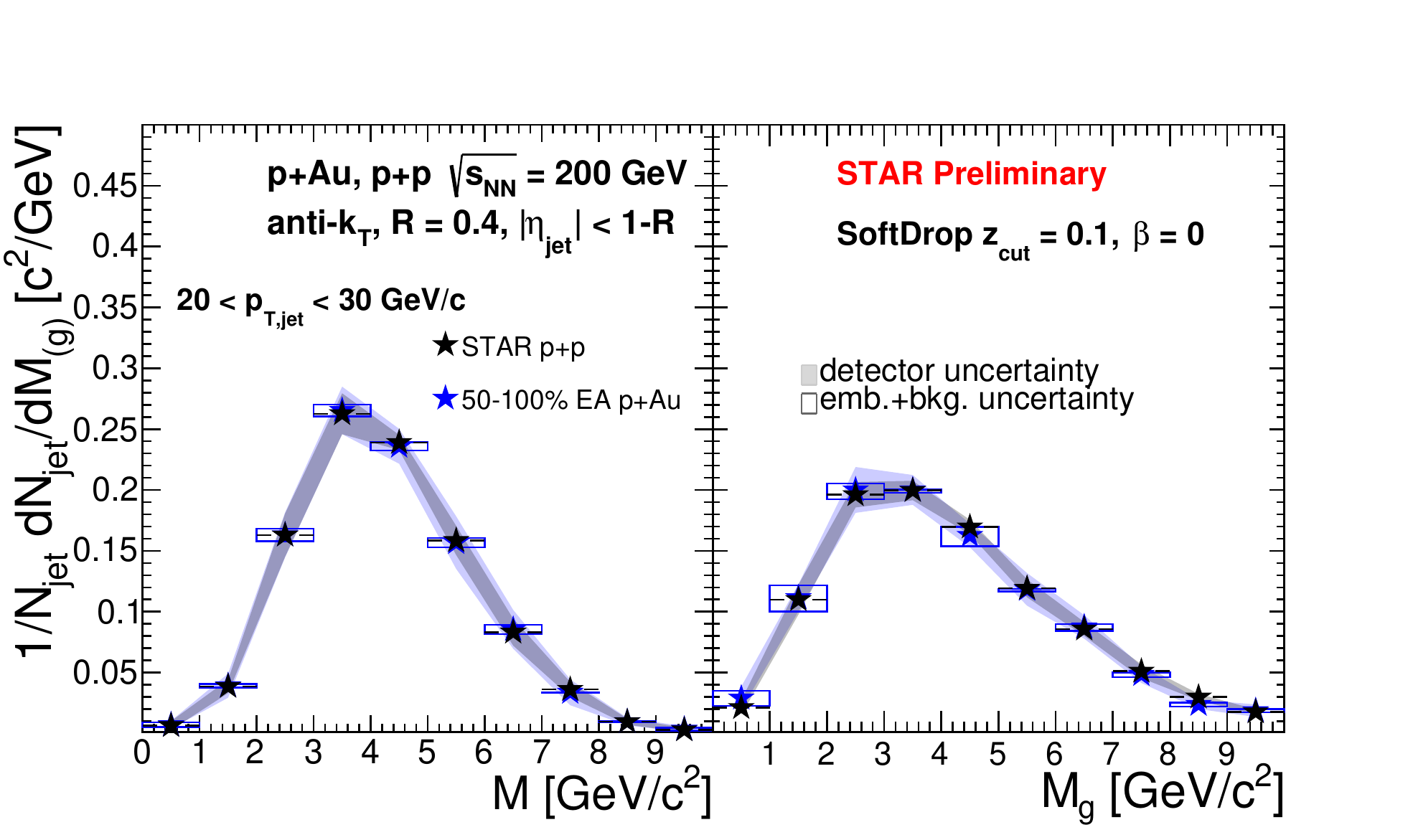}
\par\end{centering}
\caption{Measurements of jet mass, $M$ (left), and groomed jet mass, $M_{\text{g}}$
(right), in low event activity $p$+Au collisions (blue stars), compared
to those in $p$+$p$ collisions (black stars) as shown in Fig.~\ref{fig:ppmass}
for a single jet $p_{\text{T}}$ selection, $20<p_{\text{T}}<\unit[30]{GeV}/c$.
See $\S$~\ref{sec:Measurement} for a definition of event activity.
The shaded bands denote the uncertainties that are common between
the $p$+Au and $p$+$p$ analyses, while the boxes denote the additional
embedding and background uncertainty assessed for the $p$+Au data.\label{fig:pAperiphmass}}
\end{figure}
Figure~\ref{fig:pAperiphmass} shows the comparison of the fully
corrected jet mass (left) and groomed jet mass (right) in low-EA $p$+Au
collisions (blue stars) to $p$+$p$ collisions (black stars) for
jets with $20<p_{\text{T}}<\unit[30]{GeV}/c$. We observe no significant
difference between them, which is expected.

\begin{figure}
\begin{centering}
\includegraphics[width=0.69\textwidth]{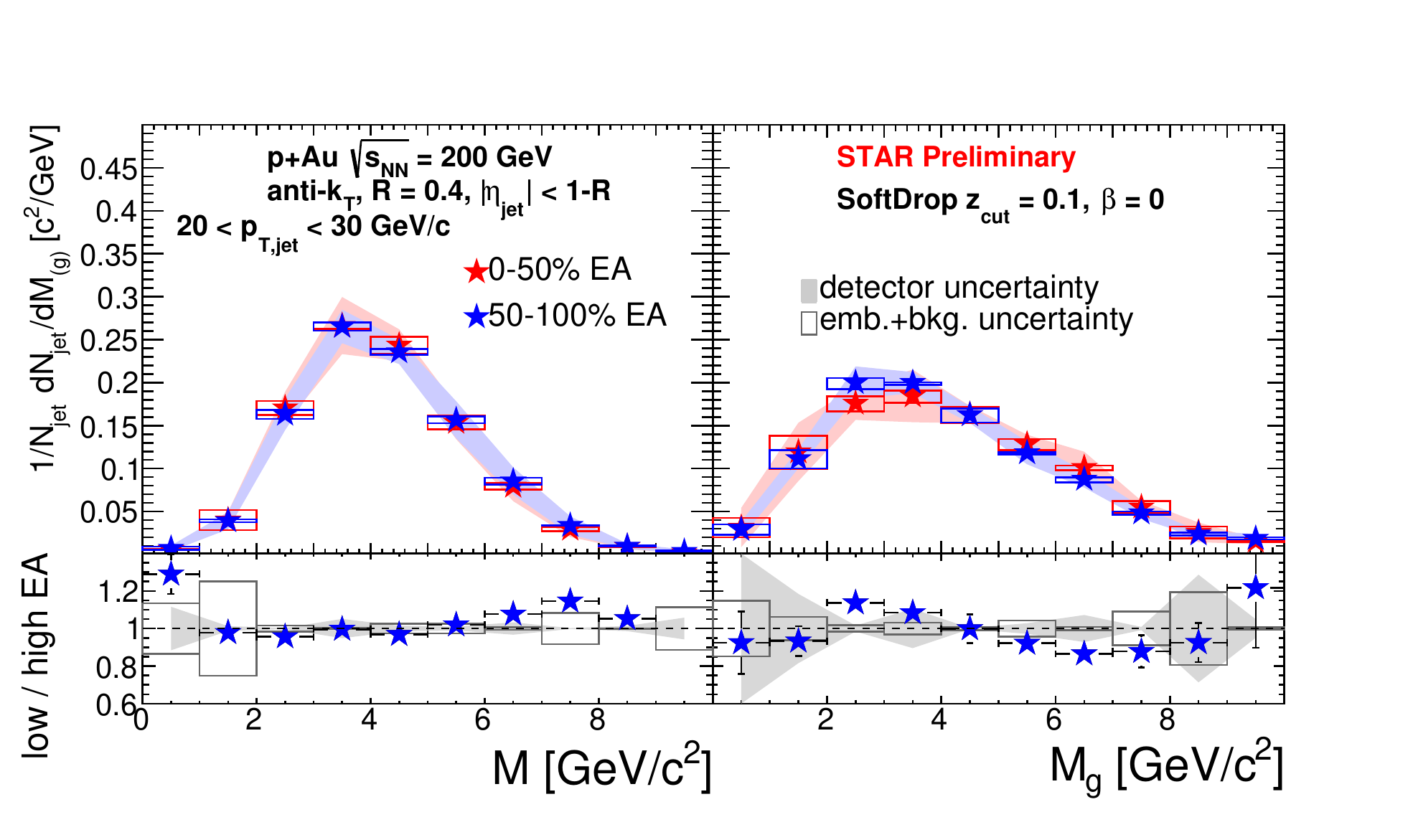}
\par\end{centering}
\caption{Measurements of ungroomed jet mass, $M$ (left), and groomed jet mass,
$M_{\text{g}}$ (right), in high-EA $p$+Au collisions (red stars),
compared to low-EA\textbf{ }$p$+Au collisions (blue stars) as shown
in Fig.~\ref{fig:pAperiphmass} for a single jet $p_{\text{T}}$
selection, $20<p_{\text{T}}<\unit[30]{GeV}/c$. See $\S$~\ref{sec:Measurement}
for a definition of event activity. The ratio between the two event
activity classes is shown in the bottom panel. See Fig.~\ref{fig:pAperiphmass}
for more details on the curves.\label{fig:pAcentmass}}
\end{figure}
In Fig.~\ref{fig:pAcentmass}, we compare the fully corrected jet
mass in low-EA $p$+Au collisions (blue stars) to high-EA $p$+Au
collisions (red stars). The jet mass between the two is consistent
within uncertainties, suggesting that the jet structure is unmodified
by cold nuclear matter effects in high-EA $p$+Au collisions. Additionally,
the groomed jet mass exhibits similar behavior, indicating that the
core of the jets is unmodified as well.

\section{Conclusions\label{sec:Conclusions}}

We have presented the first fully corrected inclusive jet mass measurements
in $p$+$p$ and $p$+Au collisions at STAR. The $p$+$p$ jet mass
measurements present an opportunity for further Monte Carlo tuning,
while the $p$+Au jet mass measurement indicates that jet substructure
is not significantly affected by CNM effects. It is possible that
competing effects on the angular and momentum scales of the jet are
cancelled in the mass, so we will investigate the groomed jet momentum
fraction, $z_{\text{g}}$, and radius, $R_{\text{g}}$, for a full
suite of jet substructure observables in $p$+Au collisions. Additionally,
jet resolution parameter dependence will be investigated, and event
activity selections narrowed to enhance potential cold nuclear matter
effects on the jet mass. Finally, we will use this measurement as
a baseline for a similar measurement in the hot nuclear environment
of Au+Au collisions.

\end{document}